\documentstyle[floats,aps,twocolumn]{revtex}
\input epsf

\begin{document}

\title{Tunneling spectroscopy in small grains of superconducting MgB$_2$}

\author{G. Rubio-Bollinger, H. Suderow $^{+}$, S. Vieira }

\address{ Laboratorio de Bajas Temperaturas*, Dept. F\'\i sica de la Materia Condensada C-III,
 Instituto Universitario de Ciencia de Materiales "Nicol\' as Cabrera",
 Universidad Aut\' onoma de Madrid, E-28049 Madrid, Spain.
 \\
 $^{+}$Intituto de Ciencia de Materiales de Madrid,
 Consejo Superior de Investigaciones Cient\'ificas,
 Campus de Cantoblanco, E-28049 Madrid, Spain.}


\maketitle

\begin{abstract}

We report on tunneling spectroscopy experiments in small grains of
the new binary intermetallic superconductor MgB$_2$. Experiments
have been performed at 2.5 K using a low temperature scanning
tunneling microscope. Good fit to the BCS model is obtained, with
a gap value of 2 meV. In the framework of this model, this value
should correspond to a surface critical temperature of 13.2 K. No
evidence of gap anisotropy has been found.\\


\end{abstract}

The recent discovery of  superconductivity at 39 K in magnesium
diboride by Nagamatsu \textit{et al.} \cite{Akimitsu}, has
produced a notable excitement in the people investigating in the
field. This novel achievement is extremely exciting as it opens a
new route towards the search of high temperature superconductors.
Indeed, the composition and the hexagonal structure of this
compound is simpler than those of the other known systems that are
superconducting at these or higher temperatures, as e.g. the
Copper-oxide High $T_c$ materials or as the C$_{60}$ based
compounds. First measurements of the thermodynamic properties
\cite{Finnemore} (magnetization and specific heat) and of the
isotope effect on the B atoms are already available \cite{Budko}.
These data demonstrate that this system is a type II
superconductor (with Ginzburg-Landau parameter $\kappa\simeq 26$)
and that replacement of $^{10}$B by $^{11}$B leads to a decrease
of the critical temperature, in agreement with phonon mediated BCS
superconductivity.

As recently reported \cite{Kortus}, band structure calculations
show a highly isotropic character for the electronic properties of
this compound, favoring the phonon mediated superconductivity.
Other very recent theoretical proposals by Hirsch \cite{Hirsch}
have pointed towards hole superconductivity, based in the negative
values for the Hall coefficient in other metal diborides with the
same crystal structure. This author proposes several experimental
tests of his theory, one of them being the tunneling
characteristic which is predicted to be asymmetric having always a
larger current for a negatively biased sample. Tunnel spectroscopy
experiments are of keen interest to shed light on the mechanism of
superconductivity in this material.

We present in this letter tunneling characteristic curves of
MgB$_2$, obtained at liquid helium temperatures, using a home-made
low temperature scanning tunneling microscope (LT-STM) with a gold
tip as counter-electrode. We have used commercially available
magnesium diboride powder (Alfa-Aesar 98\% pure).

We have done magnetization measurements of the powder and find the
same behavior as reported in Ref.\cite{Budko}, being $T_c=37.5$ K
and the transition broader than in samples prepared by solid state
reaction of pure elements \cite{Budko}. X-ray characterization of
our powder shows an extra peak at $2\theta = 36.6^o$ due to a
small magnesium content. This inclusions correspond to a 0.2\% of
the sample volume, as obtained from a more detailed analysis of
the peak intensity.

We measure the tunneling \textit{I-V} characteristics on
individual magnesium diboride grains whose preparation involved
several steps. The MgB$_2$ powder was first dispersed in high
purity acetone in an ultrasound bath and a drop of this dispersion
was deposited on the surface of a high purity gold sample. Acetone
was then evaporated putting the sample in an oven at 80 $^o$C.
Then the dry disperse powder was gently pressed into the gold
surface with a flat synthetic rubi, forcing the hard magnesium
diboride grains to penetrate into the softer gold substrate. The
sample was then immersed in an acetone ultrasound bath in order to
remove grains that were not tightly fixed to the gold substrate.
Inspection of the resulting sample with an optical microscope
showed a distribution of single MgB$_2$ grains separated by clean
gold regions several microns wide. This preparation method
overcomes difficulties arising from the preparation of pellets
that often result in artifacts in the \textit{I-V} curves possibly
due to bad intergrain connection \cite{jgr}. In addition, the
grains of MgB$_2$ result embedded in the gold matrix and are
expected to have a good electrical connection to the substrate,
and therefore to the electrodes. In Fig. 1 we present two scanning
electron microscopy (SEM) images of grains embedded in the gold
substrate.

\begin{figure}[h]
\vspace{0mm}
\begin{center}
        \leavevmode
        \epsfxsize=80mm
        \epsfbox{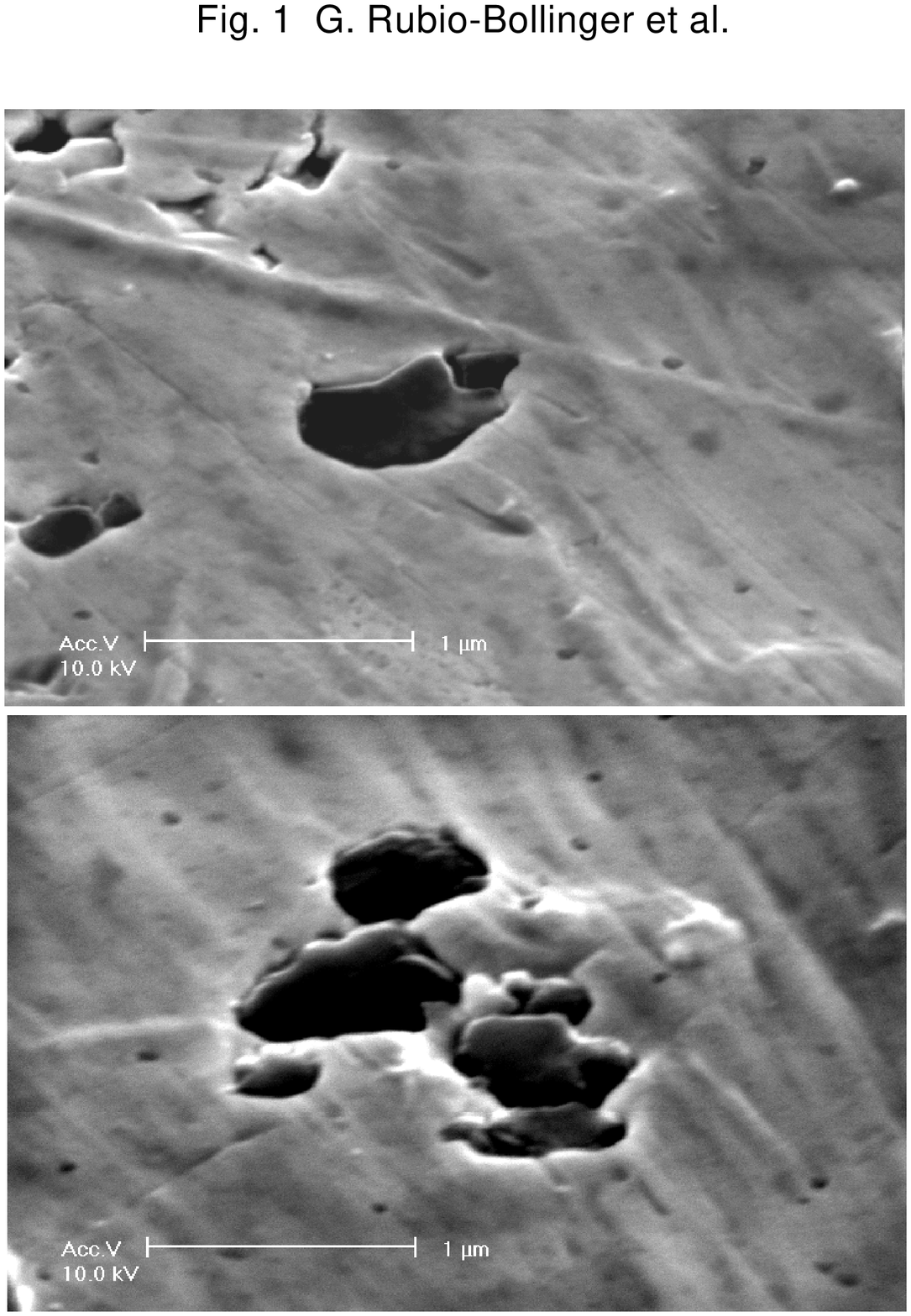}
 \end{center}
\vspace{0mm} \caption[]
 { SEM micrographs of the magnesium diboride grains (dark)
embedded in the gold substrate. Typical grain sizes are between
0.5 $\mu$m and 1 $\mu$m, thus much larger than the superconducting
coherence length. } \label{FIG. 1.}
\end{figure}

In order to be able to find on the gold substrate different
individual grains we use a home-build STM operating at low
temperature supplemented with a coarse X-Y positioning stage
\cite{MesaXY}. This stage allows for precise positioning of the
tip over the entire sample surface (2x2 mm$^2$) in steps as small
as 20 nm, without impairing its mechanical stability.
Characteristic \textit{I-V} curves in the tunneling regime were
recorded with the STM in a four terminal configuration. The
voltage and the current are measured using two low-noise
differential preamplifiers and all lines connecting the tip and
the sample to the room temperature electronics are carefully
filtered by feed-through capacitors and a combination of lossy
coaxial and twisted pair cables. This electronic setup has been
previously shown to be necessary to avoid external electromagnetic
noise to smear out superconducting features in tunneling
spectroscopy measurements with local probe techniques
\cite{scheer}\cite{suderow}.

The measured  tunneling \textit{I-V} curves obtained at the clean
gold surface are ohmic. However, using the STM coarse X-Y
positioning stage we can find locations at which the \textit{I-V}
curves show superconducting features, indicating that the tip is
on top of a MgB$_2$ grain. In addition, the topographic STM images
at these locations look very flat, probably due to the appearance
of crystalline faces. We scanned along many different,
well-separated (10 $\mu$m) positions on the surface and therefore
studied several grains. Current vs distance curves obtained at the
MgB$_2$ grains show exponential behaviour characteristic of
tunneling with a work function of about 0.5 eV.

\begin{figure}[h]
\vspace{0mm}
\begin{center}
        \leavevmode
        \epsfxsize=80mm
        \epsfbox{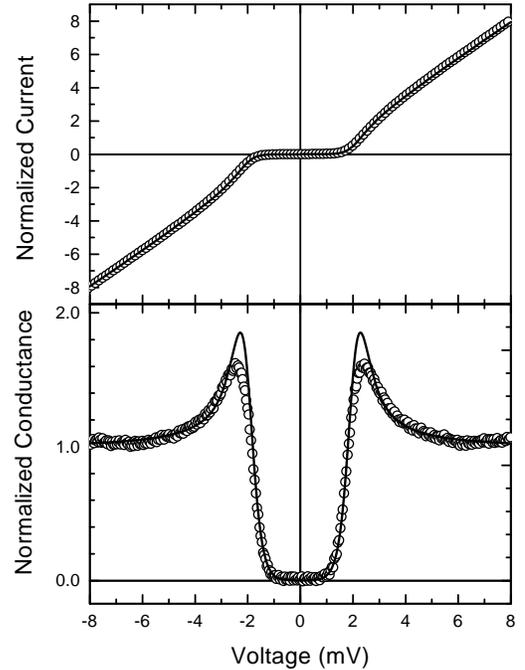}
 \end{center}
\vspace{0mm} \caption[]
 {Experimental (open circles) tunneling \textit{I-V}
characteristic curve and its differential conductance, normalized
to $R_N=2.5 $ M$\Omega$. The solid line corresponds to a BCS
normal-superconductor tunnel curve with a gap of $\Delta=2.0$ meV
and $T=2.5$ K. } \label{FIG. 2.}
\end{figure}

In Fig. 2 we show a representative tunnel \textit{I-V} curve that
is obtained at the MgB$_2$ grains. Both the \textit{I-V} curve and
its differential conductance show the expected behavior for a
tunnel junction between a normal metal and a superconductor with
BCS density of states, using a gap $\Delta=2.0$ meV and $T=2.5$ K.
This was the temperature at which the experiment was performed. We
also show in Fig. 3 a sample of several reproducible curves taken
at different grains. All of them can be fitted, as the data of
Fig. 2, with $\Delta=2.0\pm 0.1$ meV and $T=2.5$ K.

\begin{figure}[h]
\vspace{0mm}
\begin{center}
        \leavevmode
        \epsfxsize=80mm
        \epsfbox{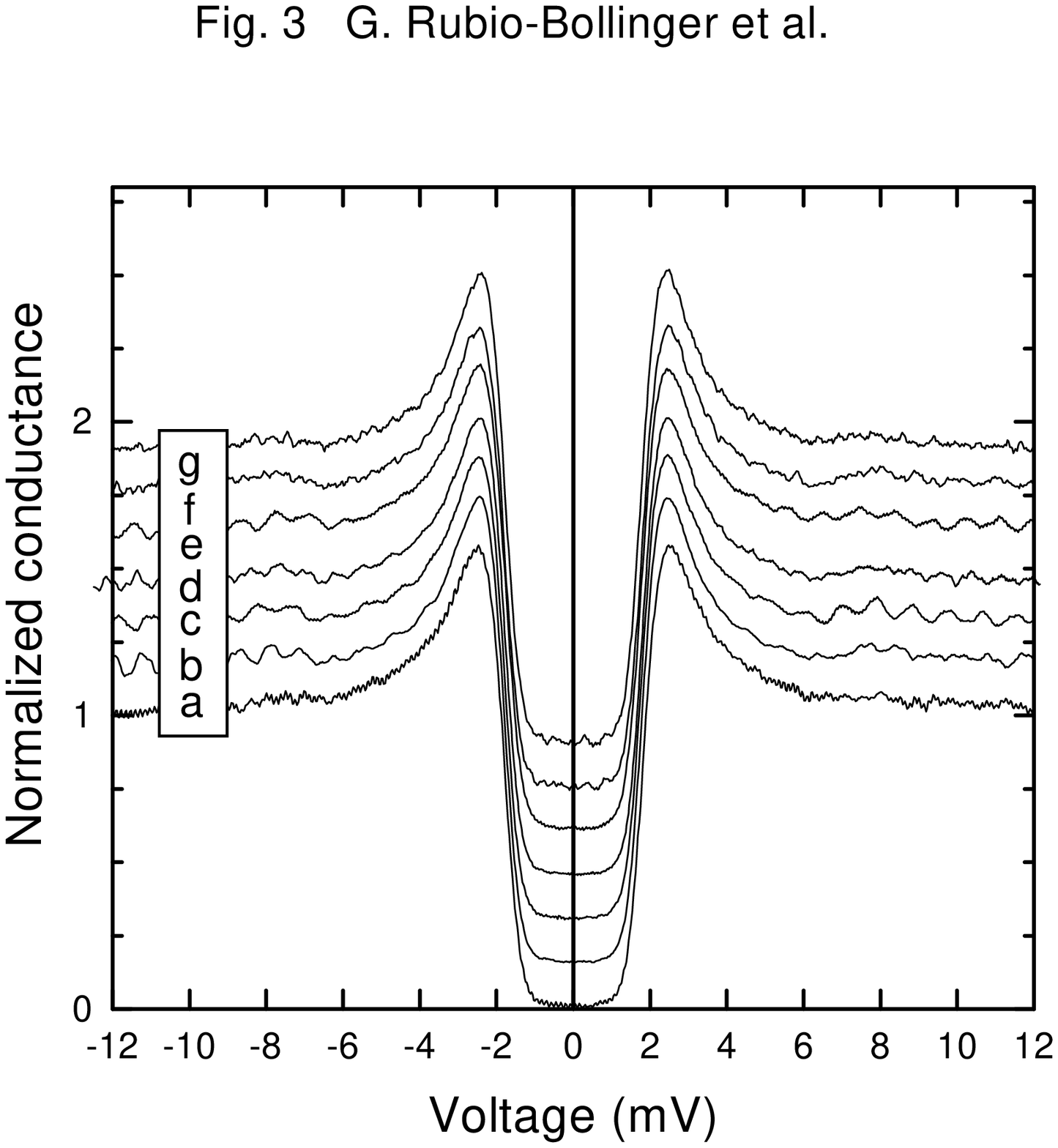}
 \end{center}
\vspace{0mm} \caption[]
 {Representative sample of several differential conductance
curves obtained at different grains. Curves a, b, c and f are
normalized to $R_N=2.5 $ M$\Omega$ and curves d, e and g to
$R_N=10$ M$\Omega$. Curves b-g are shifted for clarity by
increments of +0.15. } \label{FIG. 3.}
\end{figure}

The data show clearly that the density of states (DOS) inside the
gap is very close to zero since the curve can be fitted to a BCS
expression without any extra parameters, being thermal
quasiparticle excitations (at $T=2.5$ K) the only source of
non-zero differential conductance inside the gap. We remark that
such resolution in energy, as mentioned before, is only achievable
after proper filtering of the electric wiring of the setup.

The gap value ($\Delta=2.0\pm 0.1$ meV) for all the curves shown
in Figs. 2 and 3 is significantly lower than what would be
expected from the conventional BCS expression $\Delta=1.76k_BT_c$
(5.9 meV) taking the critical temperature that we have measured
for our powder ($T_c=$37.5 K). This discrepancy could be due to a
deviation of the DOS at the surface of this material with respect
to the bulk. Also note that in tunneling spectroscopy measurements
it is the surface DOS what is probed, which could be significantly
influenced by the presence of impurities, strong defects or
stoichiometry variations close to the surface, at distances in the
range of the superconducting coherence length $\xi_0=5.2$ nm
\cite{Finnemore}.

Such defective surfaces have very often been taken into account in
the calculation of tunneling characteristics via the introduction
in the model of pair-breaking centers which result in a reduced
life of Cooper pairs. Such kind of calculations, like the Dynes
model with a finite broadening parameter \cite{Dynes}, result in a
finite DOS inside the gap. However, our observations do not show
the presence of low-energy excitations within the gap.

Using the preparation method described above, where no
preferential grain orientation was induced, we are probably
testing different crystal directions, and no signal of gap
anisotropy was found. At some locations on the sample, curves
indicative of S-I-S inter-grain tunneling were found, being
consistent with the gap value measured on the free surface.

In conclusion, we have presented tunneling spectroscopy
experiments in the recently discovered superconducting compound
MgB$_2$. To do that we have developed a sample preparation method
that combined with the versatile positioning device of our low
temperature STM allows for studying many individual small
particles. In the framework of the BCS model, a critical
temperature at the surface ($T_c=13.2$ K) is found for the
measured gap value ($\Delta=2.0$ meV). We remark again, that we
have used commercial powder without any special cleaning or
regenerating process. That is in our opinion a good test of the
stability and goodness of the superconductivity at the grain
surface, region of paramount importance to prepare this compound
for its applications as a ceramic superconductor.

The authors acknowledge fruitful discussions with J.G. Rodrigo, F.
Guinea, P. Mart\' \i nez-Samper, N. Agra\" \i t, M. Mor\' an and
J. Gonz\' alez-Calvet. We thank specially I. Shmyt'ko for
providing X-ray characterization of the samples, and J.L. Mart\'
\i nez for the susceptibility measurements. We acknowledge
financial support from spanish DGIGyT under contract PB97-0068 and
also ESF program Vortex Matter in Superconductors at extreme
scales and conditions.

* Unidad asociada al ICMM.

\end{document}